\documentclass[10pt,twocolumn]{article}

\usepackage{cite}
\usepackage{amsmath,amssymb,amsfonts,amsthm}
\usepackage{algorithmic}
\usepackage{graphicx}
\usepackage{textcomp}
\usepackage{verbatim}
\usepackage{placeins}
\usepackage{algorithm} % algorithm environments
\usepackage{algorithmic}

\newcommand{\R}{\mathbb R}

\newtheorem{theorem}{Theorem}

\usepackage{geometry}
\begin{document}
\title{New Computational Techniques for a Faster Variation of \\BM3D Image Denoising}
\author{Toby Sanders and Sean Larkin
	\thanks{At the time of this work, both authors are employed by Lickenbrock Technologies, INC., St. Louis, MO, 63117 
		(email: SANDERTL20@gmail.com).
}}

\date{}

\markboth{Journal of \LaTeX\ Class Files,~Vol.~14, No.~8, August~2015}%
{Shell \MakeLowercase{\textit{et al.}}: Bare Demo of IEEEtran.cls for IEEE Journals}

\maketitle

\begin{abstract}
BM3D has been considered the standard for comparison in the image denoising literature for the last decade. Though it has been shown to be surpassed numerous times by alternative algorithms in terms of PSNR, the margins are very thin, and denoising is approaching a limiting point. The reason for the continued use of BM3D within the literature is due to its off-the-shelf ease-of-use in any application, which alternative \emph{improved} denoising algorithms sometimes fail to match. % Recently, denoising algorithms have been receiving increased attention due to their ability to be \emph{plugged-in} to more general image reconstruction tasks by way proximal algorithms, plug-and-play priors, and others. 
This article proposes a new variation of BM3D, which maintains its ease of use but is notably faster. This development brings us closer to real-time ease-of-use application of new state-of-the-art image reconstruction algorithms such as plug-and-play priors. 

We refer to our variation of BM3D as G-BM3D. In terms of image quality, our algorithm attains very similar denoising performance to the original algorithm. Though our algorithm is written completely in MATLAB software, it is already between 5-20 times faster than the original algorithm, and the modifications to the algorithm are such that it is expected to be significantly faster when ported to CUDA language and with more powerful GPUs. The improved processing time is achieved by two main components. The first component is a new computational strategy that achieves faster block matching, and the second is a new \emph{global} approach to the 3D wavelet filtering step that allows for significantly improved processing times on GPUs. The fast block matching strategy could also be applied to any of the vast number of nonlocal self-similarity (NSS) denoisers to improve processing times.
\end{abstract}

% \begin{IEEEkeywords}
% Image denoising, accelerated processing, high performance computing
% \end{IEEEkeywords}

% \IEEEpeerreviewmaketitle

\section{Introduction}
Image denoising has been a widely researched problem for decades and has made continued progress in recent years in part due to both nonlocal self-similarity (NSS) and \emph{learning-based} approaches.  These include methods such as block matching 3D (BM3D)\cite{dabov2007image}, NLM \cite{buades2005non}, TNRD \cite{chen2016trainable}, K-SVD \cite{elad2006image}, WNNM \cite{gu2014weighted}, and neural networks \cite{zhang2017beyond,burger2012image,zhang2018ffdnet}. These methods are believed to be getting close to the theoretical limit to what is possible in the realm of image denoising \cite{milanfar2012tour,chatterjee2009denoising,romano2017little} and have notably surpassed the capability of image denoising algorithms with explicit priors, such as total variation (TV) or soft wavelet thresholding. 

Recently, these powerful image denoisers have been receiving increased attention due to the new capability to be \emph{plugged-in} to more general image reconstruction tasks and inverse problems.  Most notably, a flexible framework known as Plug-and-Play Priors (P3) is a new method for coupling any image denoising algorithm with general inverse problems such as computed tomography (CT) and deconvolution \cite{venkatakrishnan2013plug,sreehari2016plug,sun2019online}. Although it is not clear what maximum a posteriori estimator (MAP) problem P3 solves, making it difficult to analyze, the method yields superior results to explicit MAP formulations such as TV regularization. Another similar alternative to P3 is Regurlarization by Denoising (RED) \cite{romano2017little,reehorst2018regularization}, and we believe interest in these areas will continue \cite{zhang2017learning,zhang2018residual,adler2018learned,meinhardt2017learning}. A challenge with these approaches still remains clear: the iterative algorithms necessary to solve P3 and RED require dozens or hundreds of implementations of a \emph{possibly} computationally intensive denoising algorithm (typically one evaluation of the denoiser in each iteration).

Perhaps the \emph{best} image denoiser, particularly in this setting, could be classified as one which attains \emph{both} good accuracy and speed, one that is simple to use, and one which is broadly applicable across many applications. Arguably no denoising technique is an outright winner in all of these categories. For example, TNRD is suitably accurate and fairly fast, but requires re-training across different applications and various noise levels. CNN denoisers tend to be fast and are the most accurate \cite{zhang2017beyond}. However, they require laborious re-training across various applications. This re-training is exacerbated by the need to empirically tune millions of parameters, and consequently the training process is very computationally intensive. This has even led some researchers to develop ideas for simpler network training by incorporating some NSS concepts into the network model \cite{lefkimmiatis2017non,vaksman2020lidia}. 

On the other hand, BM3D is very simple to use and is very accurate. It works across any imaging application, but like many leading denoising algorithms, it is computationally intensive. In this article we present the development of a variation of BM3D that is suitably accurate and significantly faster, while maintaining its ease of use. Our interest in BM3D over other alternative image denoisers as outlined above is two fold:
\begin{itemize}
	\item The image quality performance of BM3D is near state-of-the-art and competitive with almost any alternative. Though there are a number of denoisers that have been shown to attain slightly better peak signal-to-noise ratio (PSNR) performance, the margins are very thin \cite{gu2014weighted,chen2016trainable,zhang2017beyond}. It is believed that these methods are approaching a limit to what is possible in terms of removing i.i.d Gaussian noise from an image \cite{milanfar2012tour,chatterjee2009denoising,romano2017little}.
	
	\item BM3D works as a simple to use off-the-shelf denoiser for any noise level and any image application without training or any parameter tuning, since in a sense it is \emph{re-trained} on the fly for each unique input image. To denoise an image with BM3D, one has to only input the noisy image and an estimate of the noise level. In our opinion, it is almost always presented as a baseline for comparison in any modern image denoising article because of its ease-of-use and consistently high PSNR in any setting.
\end{itemize}

\subsection{Contributions and Related Work}
BM3D works by filtering small 3D volumes formed by matching similar blocks or patches from the image, and then aggregating the filtered image patches back into the denoised image. The intuition is that by forming the 3D matched blocks the algorithm exploits the redundancies naturally found in images. This is typically done in a two-stage procedure, where first the filtering is evaluated with a hard-wavelet thresholding. Then after the first filtering step, a second block-matching and empirical Wiener filter based on the first estimate is evaluated to improve the result moderately from the first estimation. In this article, we distinguish between these two steps and consider the option of only performing the first stage for the sake of computational time. We refer to the algorithm that only uses the first stage as BM3D1, and similarly the algorithm that utilizes both steps as BM3D2. As we will show, our variation of BM3D1 is far superior when considering the speed and accuracy, and our variation of BM3D2 is also notably faster than the original algorithm \cite{dabov2007image}.

The acceleration of BM3D is accomplished through several key factors. First, a faster and more elegant block matching scheme is developed through key observations and computational techniques. Effectively, the block matching step is reduced to a series of cross-correlations which are computed on CPU hardware with fast Fourier transforms (FFTs). Second, the wavelet filtering step of the locally matched blocks is carried out \emph{globally}, as described in Section \ref{sec: HT}. This not only improves the speed of the algorithm but also mildly improves the resulting denoised image. The 3D wavelet filtering is performed on a graphics processing unit (GPU) hardware with FFTs, which are massively accelerated compared with CPU FFTs, particularly due to the global filtering implemented in our variation. Within these components, we also implement two translation invariant shifting strategies that are related to wavelet cycle spinning \cite{coifman1995translation}. These strategies are described near the end of Section \ref{sec: HT}.  We refer to our accelerated variation of BM3D as G-BM3D, where the "G" is in reference to both the use of GPUs and the \emph{global} aspects of our algorithm. In the second step of BM3D is an improved Wiener filter estimate based on the first wavelet thresholding estimate. Our variation performs this step in essentially the same manner as the original algorithm, with the only improvement coming in the faster block matching scheme.

Our software implementation is currently in MATLAB and tested on a computer containing a Nvidia Titan Xp GPU with 10.7 Tera floating point operations per second TFLOPS capability. Future work will be to implement a CUDA language version that will further improve the processing speed. All comparisons of the speed and accuracy of our algorithm are made with the MATLAB algorithm made available by the original authors \cite{dabov2007image}.

%  Already, our MATLAB algorithm can process a $512\times 512$ image in less than half a second, which was found to be 8.6 times faster than BM3D1 and more than 18 times faster than BM3D2. In addition, our implementation obtains image quality very similar BM3D1, both quantitatively and in apparent visual quality. % Since BM3D is a highly engineered algorithm , in our attempt to reproduce the quality of the original algorithm, we discovered many nuances and changes to the parameter choices that make our variation quite different from the originally described algorithm. 

Previous works have attempted to study, reproduce, and/or implement acceleration for BM3D \cite{honzatko_accelerating_2017, sarjanoja_bm3d_2015, mahmoud_ideal_2017,davy_gpu_2020,lebrun_analysis_2012}. In these works, the authors' goal tends to be to develop CUDA and/or C code to reproduce an exact version of BM3D that operates faster because of hardware and software. As discovered in \cite{sarjanoja_bm3d_2015}, the nature of the original BM3D algorithm does not lend well to GPU speed up because of the memory transfer overhead, particularly for small image sizes. Even for very large 4-megapixel images, the speed up reported in their CUDA GPU version was a factor of around 4-5 over the original algorithm, and the algorithm was even slower than the original for images with 1 megapixel or less. The work proposed here is already 7-20 times faster than the original for 1-megapixel images\footnote{This primarily depends on whether we are comparing with BM3D1 or BM3D2}, while noting that our algorithm is written completely in MATLAB. The main reason is that instead of trying to reproduce the algorithm, we have modified the design and implementation in a way that lends towards faster evaluation, particularly with GPU hardware. 

Throughout this article, we do not describe the original BM3D algorithm in great detail, since this has been very well documented in the literature (see \cite{lebrun_analysis_2012} for a detailed analysis). We describe our new variation (G-BM3D) in Sections \ref{sec: BM} and \ref{sec: HT}, and the numerical results are given in Section \ref{sec: results}. Important summary discussion on the value that we feel this work brings to the image processing community is provided in Section \ref{sec: discussion}.

\subsection{Notation}
The notation used throughout most of the article is introduced here. The full image to be denoised is defined to have ${M_y\times M_x}$ pixels. Small reference blocks extracted from the image for block matching are of size $N\times N$. Large letters such as $Z\in \R^{M_y\times M_x}$ are used to denote the full images, while small letters, i.e. $z\in \R^{N\times N}$, are used to denote small image blocks extracted from $Z$. Local image regions that are used as local search windows for block matching are of size $M_{loc} \times M_{loc}$, where obviously $N<M_{loc}\le \min (M_y , M_x)$.

Images such as $Z \in \R^{M_y \times M_x}$ are indexed as $Z[i,j]$, for $i=0,1,\dots M_y$ and $j=0,1,\dots , M_x$. For indices falling outside of that range, it is implied that the image is periodic, i.e. 
\begin{equation}
	Z [i,j] := Z[i \, \text{mod} M_y , j \, \text{mod} M_x]. 
\end{equation}
The notation $z^{ij}\in \R^{N\times N}$ is used to denote the $N\times N$ block taken from $Z$ whose top left pixel is $Z[{i,j}]$, i.e.
\begin{equation}\label{eq: Fblock}
	z^{ij}[a,b] = Z[i+a,j+b ]	.
\end{equation}
In a similar fashion, $Z^{ij}\in \R^{M_y \times M_x}$ is used to denote the zero padded version of $z^{ij}$ given by
\begin{equation}\label{eq: Fpad}
	Z^{ij}[{a,b}] = 
	\begin{cases}
		Z[i+a,j+b], &\mbox{if} \quad 0\le a,b<N\\
		0, &\mbox{if} \quad otherwise.
	\end{cases}
\end{equation}

The two-norm notation used throughout the article is always implied as the Euclidean $\ell_2$-norm defined as
\begin{equation}
	\| Z \|_2^2 := \sum_{i=0}^{M_y-1} \sum_{j=0}^{M_x-1} |Z[i,j]|^2.
\end{equation}
Similarly, the inner product is always implied as
\begin{equation}
	\langle Z , Y \rangle := \sum_{i=0}^{M_y-1} \sum_{j=0}^{M_x-1} Z[i,j] \overline{Y}[i,j].
\end{equation}
Finally, $\star$ is used to denote the cross-correlation operator that takes two images and outputs the cross-correlated image defined by
\begin{equation}
	(Y	\star Z )[i,j]	:= \sum_{a =0}^{M_y-1} \sum_{b =0}^{M_x-1} Y[a,b ] Z[i+a, j+b].
\end{equation}
Note for computational purposes that the cross-correlation can be evaluated in Fourier domain as
\begin{equation}\label{eq: CCF}
	\mathcal F (Y\star Z) = \overline{\mathcal F (Y) } \cdot \mathcal F(Z),  
\end{equation}
where $\mathcal F$ denotes the discrete Fourier transform.

\section{New Fast Block Matching}\label{sec: BM}
In this section we describe the computational procedure for the fast block matching. 
Given a reference block $z^{ij} \in \R^{N\times N}$ taken from the image, the block matching to $z^{ij}$ is performed by finding blocks $z^{k \ell}$ such that
\begin{equation}\label{eq: dfg}
	d(z^{ij} , z^{k \ell}) = \| z^{ij} - z^{k \ell} \|_2^2 < \tau_{match},
\end{equation}
where $\tau_{match}$ is some set threshold.\footnote{As explained later in Section \ref{sec: HT}, our algorithm does not exactly use this criterion, but this does not impact the discussion here.}
To find such matching blocks, traditionally a straight-forward localized search optimization has been performed, where the search is only performed in a neighborhood near the reference block. We propose a new elegant and fast way to do the block matching, which only changes the computational procedure.

To describe the procedure, first recall the simple expansion formula of the norm in (\ref{eq: dfg}) given by
\begin{equation}\label{eq: expand}
	\| z^{ij} - z^{k \ell}\|_2^2 = \| z^{ij} \|_2^2 + \| z^{k \ell} \|_2^2 - 2\langle z^{ij} , z^{k \ell} \rangle.
\end{equation}
Using this expansion, the block distances may be obtained by computing the three terms in the right-hand side of (\ref{eq: expand}) independently and adding them together. In what follows, we show how this approach allows us to evaluate the distance between $z^{ij}$ and all possible $z^{k \ell}$ with only a single cross-correlation operation.

\begin{theorem}\label{thm: BM}
	Let $Z\in \R^{M_y\times M_x}$ be an arbitrary image, and let $I_N\in \R^{M_y\times M_x}$ be given by
	\begin{equation}\label{eq: ones}
		I_N [{i,j}] = 
		\begin{cases}
			1, &\mbox{if} \quad 0\le i,j<N\\
			0, &\mbox{if} \quad otherwise.
		\end{cases}
	\end{equation}
	Let $z^{ij}, z^{k \ell}$ and $Z^{ij}$ be defined as in (\ref{eq: Fblock}) and (\ref{eq: Fpad}) respectively. Then the squared distance between $z^{ij}$ and  $z^{k \ell}$ is given by
	\begin{equation}\label{eq: expand2}
		\begin{split}
			& \quad  \| z^{ij} - z^{k \ell}\|_2^2 =\\
			&	(I_N \star Z^2)[i,j] + 	(I_N \star Z^2)[k,\ell] - 2( Z^{ij}\star Z )[k,\ell]
		\end{split}
	\end{equation}
\end{theorem}

\begin{proof}
	First, straightforward calculations can be carried out to show the following equalities:
	\begin{equation}\label{eq: 3eqLong}
		\begin{split}
			(I_N \star Z^2)[i,j] & = \| z^{ij} \|_2^2\\
			(I_N \star Z^2)[k,\ell] & = \| z^{k \ell} \|_2^2\\
			( Z^{ij}\star Z )[k,\ell] & = \langle z^{ij} , z^{k \ell} \rangle.
		\end{split}
	\end{equation}
	Combining (\ref{eq: 3eqLong}) with (\ref{eq: expand}) completes the proof.
\end{proof}

Observe from (\ref{eq: expand2}), if we precompute and store the image $I_N\star Z^2$ into memory, then the two squared norm terms in (\ref{eq: expand}) are now available for every possible block. Then for each particular block $z^{ij}$, to get its distance from every other block in the image, only one additional cross-correlation is needed, namely, $Z^{ij}\star Z$. Following this, we form the new \emph{distance} image defined as
\begin{equation}\label{eq: Dimage}
	D^{ij} := (I_N \star Z^2)[i,j] + 	I_N \star Z^2 -2 Z^{ij}\star Z ,
\end{equation}
which by Theorem \ref{thm: BM} the block distances between $z^{ij}$ and all other blocks, i.e.
\begin{equation}
	D^{ij}[k,\ell] = \| z^{ij} - z^{k \ell}\|_2^2.
\end{equation}
Hence, locating the minimum values in $D^{ij}$ will give us indices of the blocks to match to $z^{ij}$.

Finally, in practice the cross-correlation $Z^{ij}\star Z$ that is needed for each reference block is only computed locally on an image $Z_{loc}^{ij} \in \R^{M_{loc}\times M_{loc}}$. The image $Z_{loc}^{ij}$ is formed by extracting a neighborhood image centered around $z^{ij}$. The purpose of this is to reduce the computation that must be carried out many times (once for each reference block). This practice is consistent with the original BM3D algorithm, and in our own numerical tests we found that $M_{loc}=32$ was sufficiently large to result in no loss of image quality. The pseudo code for the algorithm is given in Algorithm \ref{alg: BM}.

% To summarize, the right-hand side of (\ref{eq: expand}) is used to evaluate the block distances. The $\ell_2$ norms involved in this expression are available by a precomputation as described in Corollary \ref{cor1}. The inner-product term in this expression is computed using Proposition \ref{prop1} for each reference block in a local search window around the reference block. This term is added to the normed terms that were precomputed and then the block matching is complete by obtaining the best blocks according to (\ref{eq: dfg}). The pseudo code for this algorithm is outlined in Algorithm \ref{alg: BM}. 

\begin{algorithm}[!ht]
	\caption{Block matching computational algorithm.}
	\label{alg: BM}
	\begin{algorithmic}[1]
		\STATE{Input image $Z$ for block matching.}
		\STATE{Construct "ones" image, $I_N$, defined in (\ref{eq: ones}).}
		\STATE{Evaluate $I_N \star Z^2$ to obtain all possible $N\times N$ block norms in $Z$.}
		\FOR{every reference block $z^{ij}$}
		\STATE{Form a zero-padded image block $Z^{ij}\in \R^{M_{loc}\times M_{loc}}$ as in (\ref{eq: Fpad}).}
		\STATE{Form a local window search image $Z_{loc}^{ij} \in \R^{M_{loc}\times M_{loc}}$ from $Z$ centered around $z^{ij}$.}
		\STATE{Evaluate $Z^{ij}\star Z_{loc}^{ij}$ to obtain the inner-product of $z^{ij}$ with all possible $N\times N$ blocks in $Z_{loc}^{ij}$.}
		\STATE{From the terms made available by steps 3 and 7, form $D^{ij}$ as defined in (\ref{eq: Dimage}).}
		\STATE{Store the indices of the minimum values in $D^{ij}$, which gives the block indices to match with $z^{ij}$.}
		\ENDFOR
		\STATE{Proceed to 3D collaborative filtering stage.}
	\end{algorithmic}
\end{algorithm}

The local cross-correlations (step 7 of Algorithm \ref{alg: BM}) are the dominant computations in the block matching algorithm. They were determined to be most efficient on CPU hardware using FFTs as written in (\ref{eq: CCF}).% Once the block matching is completed for all reference blocks, the next step of the algorithm is the wavelet filtering and aggregation of the blocks. Our variation on this filtering and aggregation step is given in the next section.

% \begin{figure}
	%	\centering
	%	\includegraphics[width=.45\textwidth]{blockDistancesFinal2.jpg}
	%	\caption{Diagram of our method of computing the block distances on a \emph{mostly} noise free image for all possible matched blocks of size 32$\times$32 using convolutions. Left column: precompute all block norms by convolution of squared image with 32$\times$32 ones block. Right column: compute inner product of each possible matched block with the reference block by convolution. Bottom right: output all block distances by combining the inner product image with the block norms image using (\ref{eq: expand}).}
	%	\label{fig: blockNorms}
	% \end{figure}

\section{New Global Volume Hard Thresholding}\label{sec: HT}

In the classical version of BM3D, after a set of blocks is matched with a reference block, a 3D orthonormal wavelet transform is applied to this $N\times N \times K$ volume, where $K$ is the number of matched blocks. A hard threshold is applied to the transformed wavelet coefficients, and the wavelet transform is inverted to produce the denoised set of image blocks. These image blocks are then aggregated back into the denoised image with various weighting schemes, and the algorithm proceeds to the next reference block. % Our variation on this approach yielded moderately improved results in our development and conveniently lends toward better acceleration performance on GPU hardware, with the only drawback being greater memory requirements.

\begin{figure}[ht!]
	\centering
	\includegraphics[width=.45\textwidth]{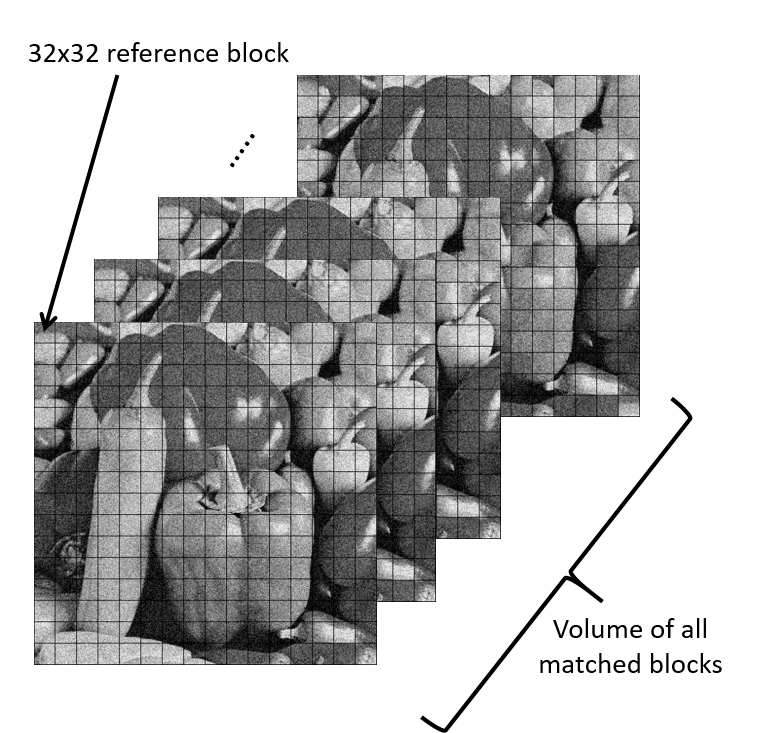}
	\caption{Diagram illustrating the 3D volume of matched blocks that are all denoised simultaneously. This demonstrates how the references blocks tile the original image, and the matched blocks for each reference block are stacked into the volume behind the noisy image.}\label{fig: Gvol}
\end{figure}

For our variation, instead of 3D filtering each set of matched blocks independently, all of the 3D matched blocks are filtered jointly as one larger volume. To accomplish this, the set of matched blocks are stacked into a larger 3D volume with spatial dimensions the same size as the image. The first slice in the $z$-coordinate of this volume contains the noisy image. The matched blocks are stacked within the $z$-coordinate slices of the volume directly behind the $x,y$-coordinates of the reference block to which they were matched (see Figure \ref{fig: Gvol}). This is repeated for each reference block. For this to work, the reference blocks are non-overlapping and tile the entire image. Moreover, the number of matched blocks is fixed to some value $K$, so that the dimension of this volume is predetermined.  After extensive testing we found $K=16$  works suitably well, in which case each reference block has exactly 15 matches.

% A minimum matching value should also be set so that every reference block has \emph{some} matches. For reference blocks that have fewer than $K$ matches, the remaining volume behind the $k$ matches is padded with the reference block to avoid unwanted edges in the volume.

Once the full volume is formed, the 3D wavelet transform is applied to the whole volume. Hard thresholding is performed on these wavelet coefficients, and then the wavelet transform is inverted to produce a denoised volume. Finally, the denoised blocks are aggregated back to produce the denoised image in almost the same way as BM3D. The reference blocks that were used to pad the volume are not included in the aggregation. 

The main advantage of our modified approach is again computational. This single filtering step on a larger volume is much more efficient than filtering each set of blocks independently. This is particularly true when considering evaluation of the algorithm on GPUs. The main computational operations involved in the wavelet transforms that perform the filtering are convolutions and hence FFTs, and the greatest performance boost with FFT evaluations on GPUs comes at larger dimensions.

% \subsection{Practical Considerations}
% In the traditional BM3D algorithm, the aggregation weights are inversely proportional to the number of nonzero wavelet coefficients remaining after the hard thresholding, with the intuition being that larger weights are used for more sparsely represented 3D matched blocks that are \emph{well-denoised}. Since we perform a global wavelet transform over the full image size, these weights are not readily available for each set of matched blocks. So a suitable surrogate is a weight which is inversely proportional to the 3D total variation of each set of matched blocks. Even a simple weight equal to 1 did not significantly diminish the results, which was also observed in \cite{lebrun_analysis_2012}.

To complete our method, some circle-shifting and averaging is used to improve the result. Where in the original algorithm the reference blocks are free to overlap, our reference blocks tile the image due to the modifications just described. So the idea of circle-shifting the image to create new reference blocks is accomplished in two ways. First, instead of just filtering the whole volume once, it is filtered a second time by translating the volume, and the two estimates are averaged. This is a well-known approach to wavelet denoising known as translation invariant cycle-spinning \cite{coifman1995translation}. No notable improvements were observed after more than two translations. Next, similar to the cycle spinning, the entire process of matching blocks, denoising the volume, and aggregating the results is repeated after translating the noisy image by a few units. Each new translation effectively creates a new set of reference blocks, and this improves the statistics of the denoising. The final denoised image is attained by averaging the results of each these repeated denoising estimates. In our empirical results, 2-3 translations were adequate. This additional translation step does not notably increase the computational time of our algorithm since each case is run in parallel.

\subsection{Formal Details of the New Global Thresholding}
This section describes the denoising and aggregation steps in full detail. The noisy image is denoted by $Z\in \R^{M_y\times M_x}$. Let the block size be $N\times N$, where for simplicity in the exposition we assume $B_y := M_y/N$ and $B_x := M_x/N$ are integers. When this is not the case in practice the image is simply padded with mirrored images values.

The reference blocks that tile the image are denoted by $z_N^{pq}$, which is the $N\times N$ block extracted from $Z$ with the top left pixel given by $Z[p\cdot N,q\cdot N]$, for $p,q=0,1,\dots, B-1$. 
Similarly, $z^{pq}$ without the subscript $N$ is defined as in (\ref{eq: Fblock}). Then the indices for the matched blocks to $z_N^{pq}$ are defined by
\begin{equation}
	S_{pq}^{K} := \arg_{K} \min_{[i,j]\in \mathcal{N}_{pq}} \| z^{ij} - z_N^{pq} \|_2^2,
\end{equation}
where the notation $\arg_K$ returns the $K$ small arguments and $\mathcal N_{pq}$ denotes the local neighborhood search window. Then the aggregated 3D volume of matched blocks, $V\in\R^{M_y\times M_x \times K}$, has the entries
\begin{equation}
	V[p\cdot N + a ,\, q\cdot N + b,\, r^{ij}] = z^{ij}[a,b] ,
\end{equation}
where $[i,j]\in S_{pq}^{K}$, $0\le a,b <N$, and $0\le r^{ij}<K$ is the third-coordinate assigned to the matched block $z^{ij}$.

The 3D orthonormal wavelet transform used for the filtering is denoted $\mathcal T_{3D}$. Then a denoised estimate of this 3D volume is given by
\begin{equation}\label{eq: BM3DHT}
	U_0 = \mathcal T_{3D}^{-1} \left( \Upsilon \left( \mathcal T_{3D}\left(V\right) \right)\right) ,
\end{equation}
where $\Upsilon$ is a hard thresholding operator given by
\begin{equation}
	\Upsilon (\alpha ) = \begin{cases}
		\alpha, & \mbox{if} \quad |\alpha| \ge \lambda\\
		0, & \mbox{if} \quad |\alpha|<\lambda,
	\end{cases}
\end{equation}
and the threshold $\lambda$ is proportional to the noise level. Additional denoised volumes are obtained by cycle-spinning and defined by
\begin{equation}
	U_h = \mathcal{S}_{-h} (T_{3D}^{-1} \left( \Upsilon \left( \mathcal T_{3D}\left(\mathcal{S}_{h}(V) )\right) \right)\right) ,
\end{equation}
where $\mathcal{S}_h$ is the operator that circularly shifts a volume by $h$ units in each dimension. Then the final denoised volume estimate is given by averaging these estimates
$$
\mathbf{U} = \frac{1}{H} \sum_{h} U_h.
$$
In practice we have used $H=2$ and $h=0,1$. Each matched block in this volume must now be aggregated back into a 2D image to obtain the desired denoised image. This process is essentially the same as the original algorithm, which we describe below for completeness. First we must introduce some additional notation.

% Define $u^{pq}$ to be the $N\times N \times K$ block extracted from $\mathbf U$ at $[p\cdot N,q\cdot N]$. In other words, $u^{pq}$ is the filtered set of matched blocks from $S_{pq}^{K}$. 

For each $[i,j] \in S_{pq}^{K}$ and every $[p,q]$ reference block, denote $u_{ij}^{pq} \in \R^{N\times N}$ to be the 2D patch in $\mathbf{U}$ associated with the $z^{ij}$ block matched to reference block $z_N^{pq}$, so that
\begin{equation}
	u_{ij}^{pq}[a,b] = \mathbf{U} [p\cdot N + a , q \cdot N + b,r^{ij}],	
\end{equation} 
for $0\le a,b, < N$.
Similarly, we denote $u^{pq}\in \R^{N\times N\times K}$ to be the set of 3D blocks in $\mathbf U$ associated with the reference block $z_N^{pq}$. 
Finally, we define the $M_x\times M_y$ zero padded version of $u_{ij}^{pq}$ to be
\begin{equation}
	U_{ij}^{pq}[i+ a, j+ b] := 
	\begin{cases}
		u_{ij}^{pq} [a,b] &\mbox{if} \quad 0\le a,b<N\\
		0 &\mbox{if} \quad otherwise,
	\end{cases}
\end{equation}
and similarly the support function for this block is defined by
\begin{equation}
	\chi_{ij} [i+ a, j+ b] := 
	\begin{cases}
		1 &\mbox{if} \quad 0\le a,b<N\\
		0 &\mbox{if} \quad otherwise.
	\end{cases}
\end{equation}

Then the denoised image estimate is a weighted average of these block estimates just as in the original algorithm, which is given by
\begin{equation}
	Y[a,b] = \frac{\sum\limits_{p=0}^{B_y-1} \sum\limits_{q=0}^{B_x - 1} \sum\limits_{[i,j] \in S_{pq}^K} w_{pq}  U_{ij}^{pq}[a,b] }{ \sum\limits_{p=0}^{B_y-1} \sum\limits_{q=0}^{B_x - 1} \sum\limits_{[i,j] \in S_{pq}^K} w_{pq} \chi_{ij}[a,b] },
\end{equation}
The weights we use are given by
$$
w_{pq} = 1/TV(u^{pq}),
$$
where $TV$ is a 3D total variation norm.

Finally this estimate is further improved by translating the image several times, repeating the whole procedure each time, and averaging all of the results. To describe this formally, we denote the entire image denoising procedure just described by a nonlinear operator $\varphi$, so that
$$
Y = \varphi (Z).
$$
Then the final denoised estimate is given by
\begin{equation}\label{eq: firstEst}
	Y_f = \frac{1}{H}\sum_{h} S_{-h}\left( \varphi(S_h (Z))  \right),
\end{equation}
where in practice we have typically used $H=3$ and  $h = 0 , N/4, \text{ and } N/2$.

\section{Second Wiener Filter Estimation} 
The second estimate is generated in almost the same manner as the original algorithm, although speed performance is gained by using our new block matching strategy. The 3D transform in the second step is a 2D DCT transform in the $x,y$-coordinates of the matched blocks and a 1D Haar wavelet transform in the $z$-coordinate. The nature of this filtering step is such that it cannot be done globally like the first step with the wavelets. The reason for this is because the basis functions in the DCT transform are not localized in the image domain, as opposed to the localized design in image and frequency domain of wavelets. Hence, filtering globally in the DCT domain would yield poor results and ringing artifacts.

The only difference in our approach from the original algorithm is that our reference blocks again tile the image as in the first estimate, which was done for computational purposes. Additional reference blocks are gained by translating the image and repeating the strategy and averaging the results, just as in the first step. Each of the translated cases are run in parallel.

The 3D transform used in this stage is performed on all sets of matched blocks simultaneously. This is done by forming 5D tensors of dimensions $N\times N \times K \times B_y \times B_x$, where in the first 3 dimensions are the matched blocks, and the fourth and fifth dimensions take us to new reference blocks that tile the image. With these 5D tensors formed, the 3D transforms can be applied to all sets of matched blocks all at once. The filtering and aggregation rules in this step match the original algorithm precisely.

\section{Results}\label{sec: results}
\begin{figure}[ht!]
	\centering
	\includegraphics[trim={1.3cm 5.5cm 1.2cm 5.0cm}, clip,width=.45\textwidth]{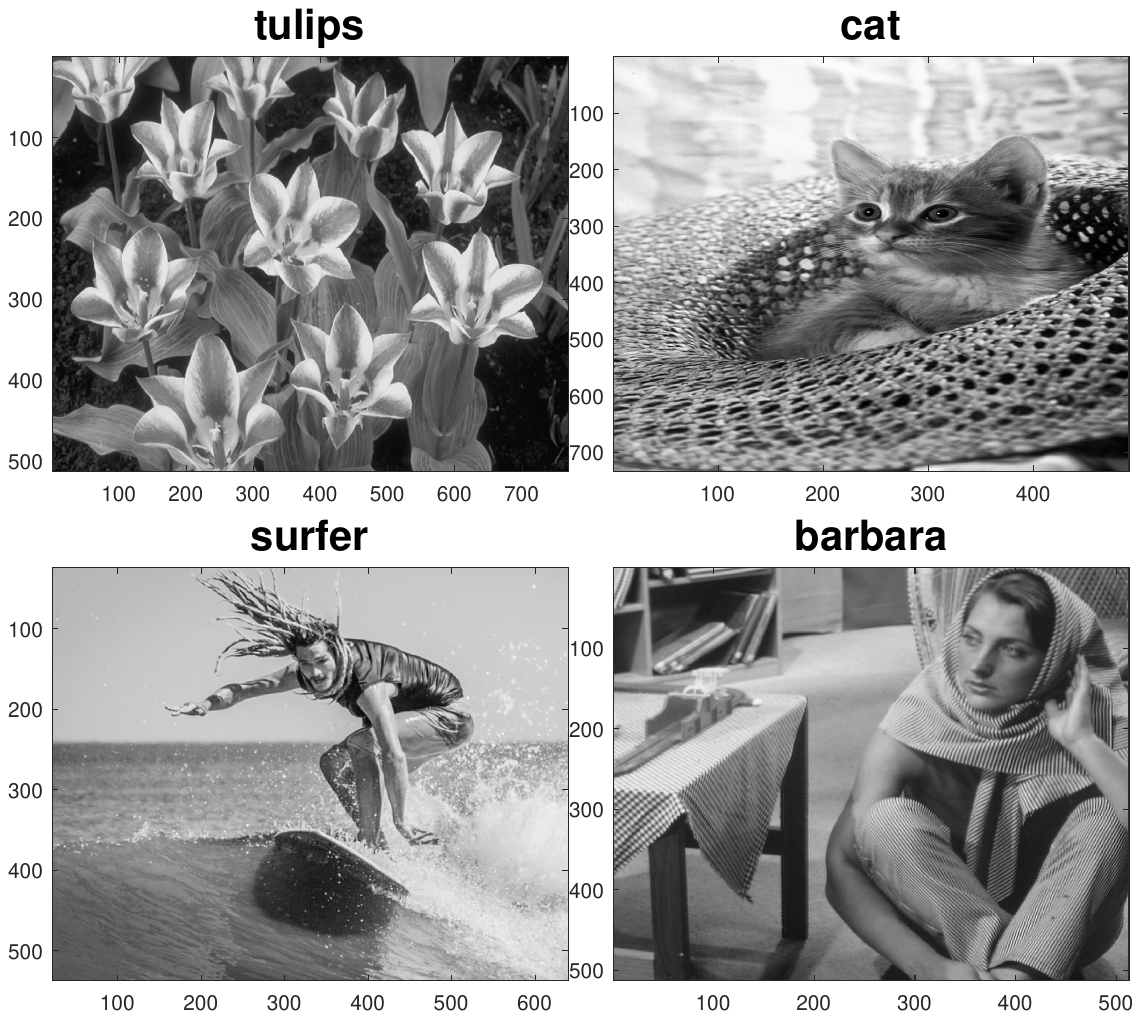}
	\caption{Some of the original test images used in our numerical simulations.}
	\label{fig: testIm}
\end{figure}

The following parameters for our algorithm were used in all numerical experiments. They were determined based on extensive prototyping and examination, empirical observations, and to strike a balance between accuracy and speed. All computations were performed in MATLAB 2020b on a Windows 10 computer containing a Nvidia Titan Xp GPU with 10.7 Tera floating point operations per second (TFLOPS) capability.
\begin{itemize}\itemsep0em
	\item Block size in first step (wavelet thresholding): $N=16$.
	\item Block size in second step (Wiener filter): $N=8$.
	\item Local search window size in cross-correlation block matching scheme: $32$.
	\item Number of matched blocks: $K = 16$.
	\item 2-D wavelet transform in spatial dimension: biorthogonal wavelets order 1.5.
	\item 1-D wavelet transform in time dimension: Haar wavelets.
	\item Number of levels in wavelet transform: 3.
	\item Number of cycles in the translation invariant wavelet denoising: 2.
	\item Number of repeated trials in the algorithm to create new reference blocks that are later averaged: 2.
	\item Hard wavelet thresholding constant: $\tau_\ell = \sigma (3.6 - 0.3*\ell)$, where $\ell = 1 , 2 ,3$ are the wavelet levels, and $\sigma$ is the standard deviation of the noise.
\end{itemize}

\subsection{PSNR Comparisons}
\begin{table}[ht]
	\centering
	\begin{tabular}{|r|c c | c c |}
		\hline 
		\textbf{PSNR} & G-BM3D1 & BM3D1 & G-BM3D2 & BM3D2 \\ \hline 
		monarch & \textbf{31.102} & 30.876 & 31.416 & \textbf{31.581} \\
		peppers & \textbf{30.731} & 30.513 & 31.112 & \textbf{31.178}\\
		tulips & \textbf{30.073} & 29.890 & 30.538 & \textbf{30.685} \\
		Lena & \textbf{30.473} & 30.175  & 30.926 & \textbf{31.080} \\
		baboon & \textbf{23.417} & 23.400  & 23.811 & \textbf{24.297} \\
		Barbara & 27.435  & \textbf{28.886} &  28.855 & \textbf{29.896}  \\
		cat & \textbf{26.527} &  26.252 & 27.168 & \textbf{27.248} \\
		surfer & \textbf{28.027}  & 27.827  & 28.213  & \textbf{28.391} \\
		
		\hline 
	\end{tabular}
	\caption{\label{table: PSNR}PSNR results after denoising with different methods. Gaussian white noise added to the image with the variance chosen so that the SNR in the noisy image is 4.}
\end{table}
Our G-BM3D algorithm was compared with BM3D on 8 different test images (see Figure \ref{fig: testIm} for examples). All comparisons of the speed and accuracy of our algorithm are made with the MATLAB algorithm made available by the original authors \cite{dabov2007image}. The noise added to the test images prior to denoising was mean zero i.i.d. Gaussian white noise with different standard deviations, $\sigma$. The values of $\sigma$ are set so that the SNR in the image is fixed at values of SNR $= 1,2,4,6,8,$ and $10$. Given a fixed value for the SNR and an abitrary image $I$, then $\sigma$ is chosen so that
$$
{SNR}  = \frac{\text{mean}(I)}{\sigma}.
$$

\begin{figure}
	\centering
	\includegraphics[width=.45\textwidth]{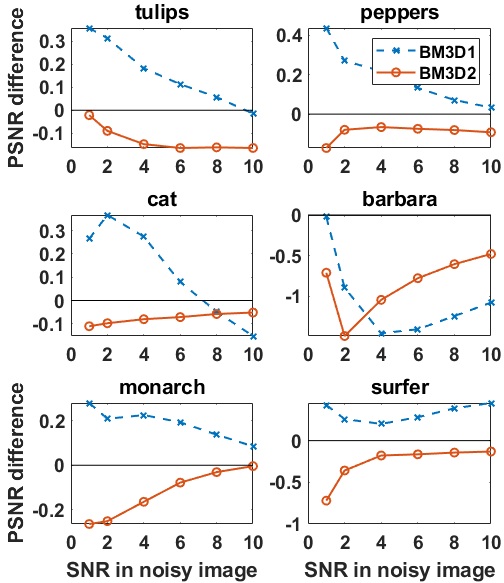}
	\caption{PSNR difference (G-BM3D minus BM3D) after denoising 6 of the 8 test images with each algorithm as a function of the SNR in the noisy image. Most cases result in a PSNR difference less than 0.5, which is subjectively is very small. The first step of our algorithm is typically more accurate (blue curves), while also providing the most speed up benefits. The second step is typically less accurate (red curves), for unknown reasons.}
	\label{fig: PSNR}
\end{figure}

The resulting PSNRs after denoising with each algorithm for the case SNR $=4$ are listed in Table \ref{table: PSNR}. Shown in Figure \ref{fig: PSNR} are PSNR comparisons for 6 of the 8 test images. Here the PSNR resulting from the original BM3D algorithm is subtracted from the PSNR resulting from G-BM3D, hence values in the plots greater than zero indicate our algorithm is better and visa-versa. Most cases result in a PSNR difference less than 0.5, which is subjectively is very small. The first step of our algorithm is typically more accurate (blue curves), while also providing the most speed improvements as shown in the next section. The second step is typically less accurate (red curves), though this variation of the algorithm was written with the intention of matching the original algorithm exactly. Therefore future tune-ups are expected to match or surpass the original algorithm, which has already been very finely tuned.

\begin{figure*}
	\centering
	\includegraphics[width=1\textwidth]{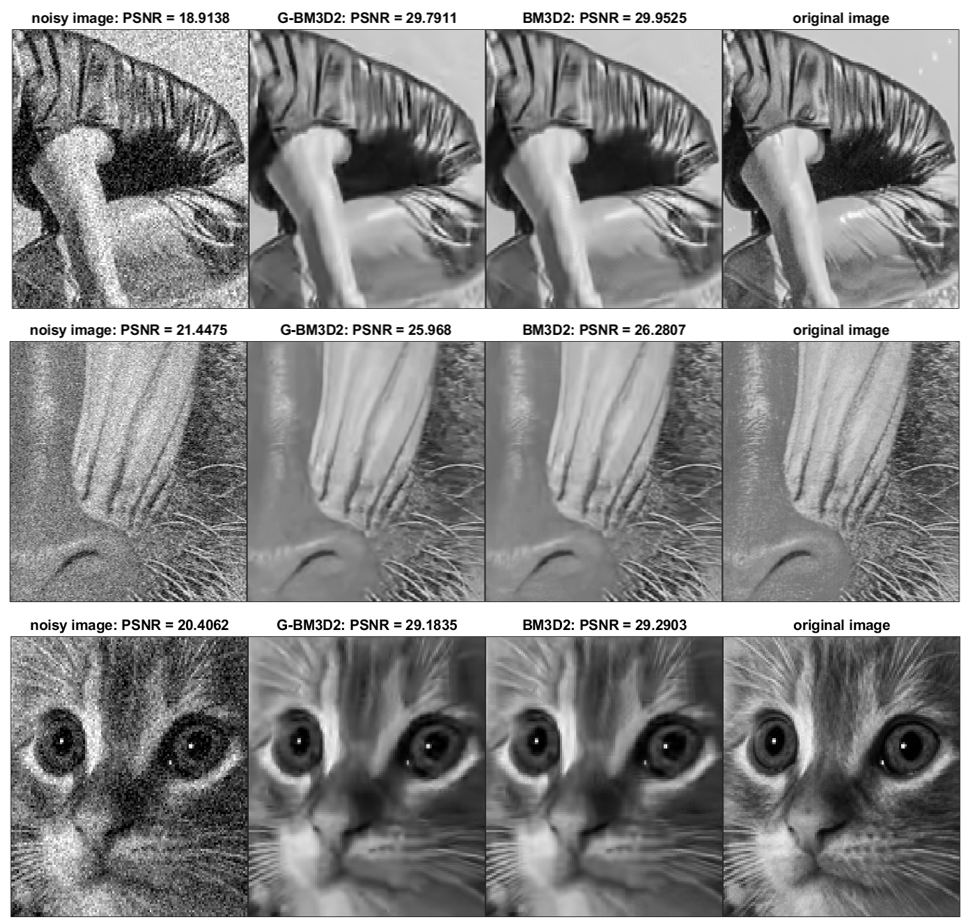}
	\caption{Image examples resulting from denoising with the different approaches, where the SNR in each noisy image is 6.}
	\label{fig: examples}
\end{figure*}

The one peculiar case is the \emph{Barbara} image, in which the original BM3D notably outperforms our algorithm. Unfortunately this seemed to be the case no matter what set of parameters we chose for our algorithm. Figure \ref{fig: examples} shows some of the images resulting from these denoising simulations, where the SNR in each of these noisy images is 6. In terms of perceived visual quality of the denoised images, we observe almost no difference between the two methods. 

\begin{comment}
	\begin{table*}[ht]
		\centering
		\begin{tabular}{|r|c|c|c|c|c|c|c|c|}
			\hline 
			\textbf{PSNR results}&  monarch & peppers & tulips & Lena & baboon & Barbara & cat & surfer \\ \hline
			G-BM3D1 & 31.102 & 30.731 & 30.073 & 30.473 & 23.417 & 27.435 & 26.527 & 28.027 \\
			BM3D1 &   30.876 & 30.513 & 29.890 & 30.175  & 23.400 & 28.886 & 26.252 & 27.827\\
			\hline  
			G-BM3D2 & 31.416 & 31.112 & 30.538 & 30.926  & 23.811 & 28.855 & 27.168 & 28.213 \\
			BM3D2 &  31.581 & 31.178 & 30.685 & 31.080 & 24.297 & 29.896 & 27.248 & 28.391\\
			\hline 
		\end{tabular}
		\caption{\label{table: PSNR}PSNR results after denoising with different methods. Gaussian white noise added to the image with the variance chosen so that the SNR in the noisy image is 4.}
	\end{table*}
\end{comment}

\subsection{Execution time}

\begin{table}[ht]
	\centering
	\begin{tabular}{r|c|c|c|c|c|c|c|}
		\textbf{Image dim.} &	$256^2$ & $512^2$ & $768^2$ & $1024^2$ & $2048^2$ & $4096^2$ \\ \hline
		G-BM3D1 & 0.170   & 0.334 & 0.532  & 0.975 & 3.491  & 14.08 \\
		BM3D1 &  1.179   & 4.257 & 8.515  & 16.92 & 71.36 & 275.5 \\ 
		G-BM3D2 & 0.787  & 1.708 & 2.861  & 5.622 & 20.58  & 83.14 \\
		BM3D2 &  2.210  & 7.985 & 19.26   & 34.50 & 159.8  & 559.8 \\
	\end{tabular}
	\caption{\label{table: speed}Execution time in seconds of the different algorithms.}
\end{table}

\begin{figure}[ht!]
	\centering
	\includegraphics[width=0.45\textwidth]{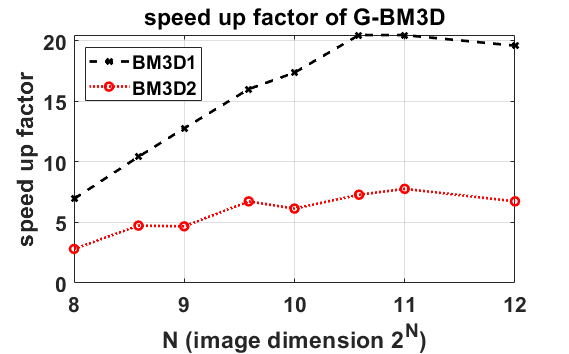}
	\caption{Comparisons of the execution times of the different denoising algorithms as a function of the image size. Plotted is the speed up factor observed from our algorithm over the original. The biggest speed gains come in performing only the first step of the algorithm (BM3D1).}\label{fig: speed}
\end{figure}

To speed up our algorithm further than the methodology already described, each large image is decomposed into smaller image patches with overlapping support and processed in parallel. These image patches are then all stitched back together after processing. The image stitching is completed after both of the two steps. The empirically determined smaller image patch size was $256 + N$, where $N$ is the block size.

We tested the execution time of each algorithm for different image sizes. All computations were performed in MATLAB 2020b on a Windows 10 computer containing a Nvidia Titan Xp GPU with 10.7 TFLOPS capability. A plot of the speed up of our G-BM3D variation as a function of the image size is shown in Figure \ref{fig: speed}. The run time of each algorithm for the different image sizes is reported in Table \ref{table: speed}.

Observe that the speed up in the first step is significant, up to 20.4 times faster than the original algorithm. The speed up is less significant for small image sizes, but the run time for these cases is already relatively fast in both cases. The speed up observed from our two-stage algorithm is up to 7.7 times faster than the original algorithm, which is less significant than BM3D1 since the 3D transforms in the second step cannot be performed globally as in the first. However, this speed up is still notable, since for example on a $2048^2$ image BM3D2 requires 159 seconds, while our algorithm only requires 21 seconds.

\FloatBarrier

\section{Discussion}\label{sec: discussion}
BM3D is designed as a two stage process, in which a first estimate is generated (BM3D1) and then used to generate a second estimate that is moderately improved (BM3D2). The first estimate is already a very well denoised image, but the second estimate makes the algorithm truly state-of-the-art. If computational time is paramount, then performing only first estimate may be an obvious compromise. % The interest in BM3D over other fast image denoisers, such as CNNs, is BM3D's ease of use, and in particular its off-the-shelf capability to work very well in any application with no parameter tuning.

We have shown the biggest advantage of our new approach comes in the first step for two main reasons. First, we have demonstrated that the speed up attained from our algorithm in the first step is overwhelming compared with the first step of the original algorithm, up to over 20 times faster. Second, as shown in the PSNR results, our algorithm also produces a more accurate first estimate. This makes the compromise of using our algorithm to only perform the first step of BM3D even more appealing. Moreover, there is seemingly no reason not to use our first step with any existing version of BM3D.

The speed up of our algorithm is attained through a new modification of the wavelet thresholding and a new computational strategy for the block-matching. The block-matching technique does not change the design of the algorithm, but only the operational procedure via cross-correlation operations. For this reason, we also propose that any block-matching algorithm should also adopt this strategy. 

The second stage of our algorithm is slightly less accurate than the original algorithm, for reasons currently unknown to us. However, in our experience in developing the algorithm, numerous incremental improvements were made along the way to get it to its current state. Therefore, it is our belief that future incremental improvements will lead us to matching or possibly surpassing the second stage of the original algorithm. As shown in the examples in Figure \ref{fig: examples}, the slightly worse PSNR in our algorithm is subjectively barely visible to the human eye. We have also shown that our two-step algorithm is up to over 7 times faster than the original algorithm, which may be a reasonable trade-off for a miniscule reduction in PSNR. For all of these reasons, we feel our two-step variation should be very appealing to practitioners.

We also note that the biggest speed gains in our algorithm come for larger image sizes, where the computational burden becomes significantly greater making speed more important. Let us consider for example a potential use case where our algorithm may be preferred. Suppose a user is denoising a $4k\times 4k$ image (a typical image size for today's cameras) and processing time is important, then they may choose to use G-BM3D1, which requires only 14 seconds, while BM3D2 requires over 9 minutes. This development also brings us closer to real time application of plug-and-play prior like algorithms, that use BM3D or other denoising algorithms as a major component in image reconstruction.

\section*{Acknowledgment}
The authors would like to thank Brendt Wohlberg for his helpful suggestions.

% \bibliography{my_bib}
% \bibliographystyle{abbrv}

\end{document}